\documentclass[aps,prl,twocolumn,groupedaddress,showpacs]{revtex4}
\usepackage{bm}
\usepackage{graphicx}

\begin{document}

\title{Spin Polarization of Gapped Dirac Surface States Near the Topological Phase Transition in TlBi(S$_{1-x}$Se$_{x}$)$_{2}$}
\author{S. Souma,$^1$ M. Komatsu,$^2$ M. Nomura,$^2$ T. Sato,$^2$ A. Takayama,$^2$ T. Takahashi,$^{1,2}$ K. Eto,$^3$ Kouji Segawa,$^3$ and Yoichi Ando$^3$}
\affiliation{$^1$WPI Research Center, Advanced Institute for Materials Research, 
Tohoku University, Sendai 980-8577, Japan}
\affiliation{$^2$Department of Physics, Tohoku University, Sendai 980-8578, Japan}
\affiliation{$^3$Institute of Scientific and Industrial Research, Osaka University, Ibaraki, Osaka 567-0047, Japan}

\date{\today}

\begin{abstract}
We performed systematic spin- and angle-resolved photoemission spectroscopy of TlBi(S$_{1-x}$Se$_{x}$)$_{2}$ which undergoes a topological phase transition at $x$ $\sim$  0.5.  In TlBiSe$_2$ ($x$ = 1.0), we revealed a helical spin texture of Dirac-cone surface states with an intrinsic in-plane spin polarization of $\sim$ 0.8.  The spin polarization still survives in the gapped surface states at $x$ $>$ 0.5, although it gradually weakens upon approaching $x$ = 0.5 and vanishes in the non-topological phase.  No evidence for the out-of-plane spin polarization was found irrespective of $x$ and momentum.  The present results unambiguously indicate the topological origin of the gapped Dirac surface states, and also impose a constraint on models to explain the origin of mass acquisition of Dirac fermions.
\end{abstract}
\pacs{73.20.-r, 71.20.-b, 75.70.Tj, 79.60.-i}

\maketitle

  Three-dimensional topological insulators (TIs) exhibit a novel quantum state with metallic topological surface state (SS) which disperses across the bulk band gap generated by a strong spin-orbit coupling.  Specifically, the topological SS is characterized by a linearly dispersing Dirac-cone energy band with the helical spin texture \cite{HasanRMP, SCZhangReview, XiaNP, SoumaZspinPRL}, which hosts massless Dirac fermions protected by time-reveal symmetry (TRS).  This peculiar SS of TIs provides a platform for fascinating quantum phenomena such as the robustness against nonmagnetic impurities / disorder \cite{TaskinMBE, SeoNature} and the emergence of Majorana fermions \cite{Majorana}.  Breaking the TRS by a magnetic field or a magnetic order lifts the band degeneracy at the Dirac point and causes an energy gap called a Dirac gap in the topological SS, turning the massless Dirac fermions into a massive state.  Theoretically, realization of the massive state is a prerequisite to novel topological phenomena such as the topological magnetoelectric effect and half-integer quantum Hall effect \cite{QiMMP, ME}, but experimentally, the role of TRS-breaking magnetic impurities in triggering the Dirac-gap opening in the topological SS is under intense debate \cite{ShenScience, Valla, Rader, HasanFe}.

Recently, it was found by angle-resolved photoemission spectroscopy (ARPES) measurements that the thallium-based ternary chalcogenide solid-solution TlBi(S$_{1-x}$Se$_{x}$)$_{2}$  goes through a quantum phase transition (QPT) from the topological ($x$ $>$ 0.5) to the non-topological ($x$ $<$ 0.5) phase \cite{HasanTBSScience, SatoNP}.  Intriguingly, the Dirac-cone topological SS in TlBiSe$_2$ ($x$ = 1.0) starts to show a Dirac gap upon partially replacing Se with S, and the gap gradually grows upon increasing the S content until the QPT at $x$ = 0.5 eliminates the SS \cite{SatoNP}.  Since there is no evidence for magnetic impurities in the measured crystals, it has been puzzling how the Dirac fermions in TlBi(S$_{1-x}$Se$_{x}$)$_{2}$ acquire the mass without explicitly breaking the TRS.  To gain insights into the microscopic mechanism of this intriguing mass acquisition, it is of particular importance to establish the spin structure of the gapped SS in TlBi(S$_{1-x}$Se$_{x}$)$_{2}$, since the topological or non-topological nature of the SS as well as the origin of the Dirac gap is intimately related to the characteristics of the spin texture.

In this Letter, we report our spin-resolved ARPES results of TlBi(S$_{1-x}$Se$_{x}$)$_{2}$ with various $x$ values ($x$ = 1.0, 0.9, 0.8, 0.6, and 0.0), and present the evolution of the spin texture as a function of $x$.  We demonstrate that the helical spin texture persists even in the gapped SS observed at $x$ $>$ 0.5, while the spin polarization vanishes in the ARPES spectrum in the non-topological phase at $x$ $<$ 0.5 in accordance with the absence of the SS.  In addition, we found unusual reduction of the spin polarization in the SS near the QPT.  We discuss the implications of the present experimental findings on the mechanism of the Dirac gap opening.
    
    High-quality single crystals of TlBi(S$_{1-x}$Se$_{x}$)$_{2}$ were grown by a modified Bridgman method.  Details of the sample preparations were described elsewhere \cite{SatoNP}.  ARPES measurements were performed with a spin-resolved photoemission spectrometer based on MBS-A1 analyzer at Tohoku University \cite{SoumaSpinRSI}.  We used one of the Xe I lines ($h\nu$ = 8.437 eV) to excite photoelectrons. Samples were cleaved $in$ $situ$ along the (111) crystal plane in an ultrahigh vacuum of 5$\times$10$^{-11}$ Torr.  The energy resolution for the spin-resolved and regular ARPES measurements was set at 40 and 6 meV, respectively.  The sample temperature was kept at 30 K during the measurements.  We used the Sherman function value of 0.07 to obtain spin-resolved ARPES data \cite{SoumaSpinRSI}.
   
   \begin{figure}[t]
\includegraphics[width=3.4in]{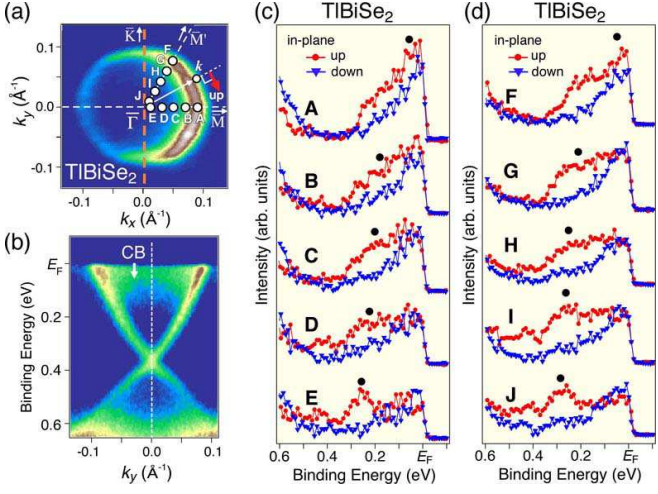}
\vspace{-0.4cm}
\caption{(Color online) (a) ARPES intensity at $E_{\rm F}$ around the $\bar{\Gamma}$ point for TlBiSe$_2$ plotted as a function of two-dimensional wave vector {\bf k} measured with the Xe I resonance line ($h\nu$ = 8.437 eV) at $T$ = 30 K.  The ARPES intensity is integrated within $\pm$20 meV with respect to $E_{\rm F}$.  The definition of an in-plane up-spin vector is also indicated by an arrow.  (b) Near-$E_{\rm F}$ ARPES intensity along the $\bar{\Gamma}$$\bar{K}$ cut for TlBiSe$_2$.  (c, d) Spin-resolved EDCs for the {\bf k} points (A-E and F-J) whose positions in the 2D Brillouin zone are indicated by open circles in (a).
}
\end{figure}
   
 First we demonstrate the spin structure of TlBiSe$_2$ ($x$ = 1.0) in which the topological origin of the Dirac-cone SS is well established by previous studies \cite{TBSSatoPRL, TBSShenPRL, TBSKurodaPRL}.  Figure 1(a) shows the ARPES intensity plot at $E_{\rm F}$ of TlBiSe$_2$ around the $\bar{\Gamma}$ point as a function of in-plane wave vector $k_x$ and $k_y$.  We clearly observe circular Fermi surface (FS) with hexagonal deformation, which originates from the upper branch of the Dirac-cone SS as displayed in Fig. 1(b).   We also find a finite intensity distribution around the $\bar{\Gamma}$ point as indicated by white arrow in Fig. 1(b).  This signal is attributed to the bulk conduction band (CB) which is slightly occupied due to electron doping caused by Se vacancies in the crystals \cite{TBSSatoPRL, TBSShenPRL, TBSKurodaPRL}.
  
 Figures 1(c) and (d) show spin-resolved energy distribution curves (EDCs) of TlBiSe$_2$ for the in-plane spin component measured at various  {\bf k} points along the $\bar{\Gamma}$$\bar{M}$ cut as indicated in Fig. 1(a).  We define the spin polarization vector perpendicular to the measured {\bf k} [see, $e.g.$, thick red arrow in Fig. 1(a)], where the ``up spin'' points to the clockwise direction.  One can immediately recognize that the up-spin EDC is always more prominent than the down-spin counterpart in all the measured {\bf k} region, consistent with the spin helical texture of the Dirac-cone SS \cite{SoumaZspinPRL, HasanTBSScience}.  Upon closer look, one finds that the shape of the up-spin EDC depends on the {\bf k} point.  For instance, near the $\bar{\Gamma}$ point (point E and J) the peak in the up-spin EDC (marked by circle) is located at $\sim$ 0.3 eV, but it approaches $E_{\rm F}$ on moving away from the $\bar{\Gamma}$ point (from E to A, or from J to F), reflecting an electronlike dispersion of the upper branch of the Dirac cone (note that the dispersive feature is less clear than the regular ARPES data because of the wider {\bf k} window and the lower energy resolution of the spin-resolved mode).  On the other hand, the peak in the down-spin EDC is always pinned at $E_{\rm F}$.  This is because it is governed by the bulk CB which contributes equally to the up- and down-spin EDCs around the  $\bar{\Gamma}$ point (as evidenced by the small difference in the near-$E_{\rm F}$ spectral weight between the up- and down-spin data at points D, E, I, and J) as a result of the spin-degenerated character.
 
 \begin{figure}[t]
\includegraphics[width=3.4in]{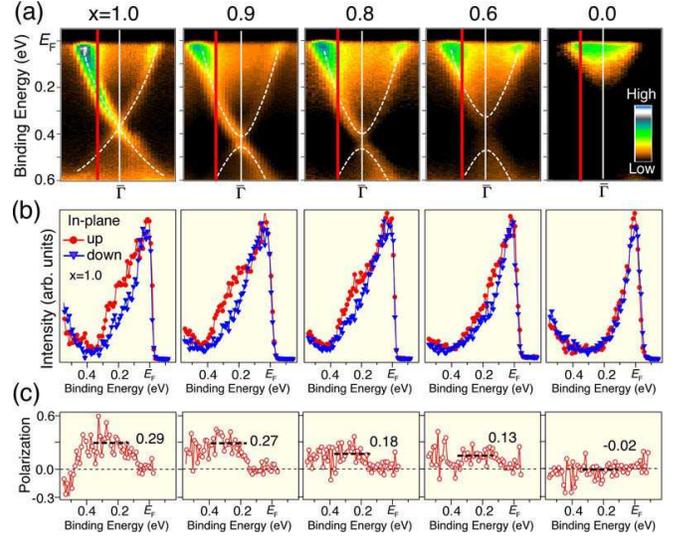}
\vspace{-0.4cm}
\caption{(Color online) (a) $x$-dependence of near-$E_{\rm F}$ ARPES intensity around the $\bar{\Gamma}$ point for TlBi(S$_{1-x}$Se$_{x}$)$_{2}$.  Dashed lines are guides to the eyes to trace the band dispersion.  The vertical red (thick) line represents the  {\bf k} point where the spin-resolved EDCs were obtained.  (b, c) Corresponding spin-resolved EDCs for the specified  {\bf k} point and their spin polarization $P_{\rm IP}$, respectively.  In (c), the averaged $P_{\rm IP}$ value in the energy window of $\pm$0.1 eV with respect to the energy of the SS ($E_{\rm SS}$) is indicated by a number and also by a dashed line.  The experimental uncertainty (error bar) in the averaged $P_{\rm IP}$ value is $\pm$0.05.}
\end{figure}
 
Figure 2(a) shows the near-$E_{\rm F}$ ARPES intensity plot of TlBi(S$_{1-x}$Se$_{x}$)$_{2}$ measured along the $\bar{\Gamma}$$\bar{M}$ direction ($k_x$) for various $x$ values ($x$ = 1.0-0.6 in the topological phase and $x$ = 0.0 in the non-topological phase).  The ``X''-shaped topological SS transforms into the gapped SS in $x$ = 0.9-0.6, as one can see from the suppression of the ARPES intensity at the Dirac point and the parabolic-like dispersion of the upper branch of the Dirac cone.  The characteristics of the gapped SS and its intrinsic nature are demonstrated in Supplemental Materials, where it is explicitly shown that the Dirac gap opening in our data is not due to a spectral-weight suppression at the Dirac point, momentum broadening, bad cleaves, nor a random disorder effect. At $x$ = 0.0 (TlBiS$_2$), the SS completely disappears and only the bulk CB is observed at $E_{\rm F}$, reflecting the non-topological nature \cite{SatoNP}.
 
 To elucidate the evolution of the spin structure of the SS, we have performed spin-resolved ARPES measurements by focusing on a particular  {\bf k} point.   The  {\bf k} point was chosen to be located midway between the Fermi vector ($k_{\rm F}$) and the  $\bar{\Gamma}$ point along the $\bar{\Gamma}$$\bar{M}$ direction [see vertical red (thick) line in Fig. 2(a)] so that the binding energy ($E_{\rm B}$) of the surface band ($E_{\rm SS}$ $\sim$ 0.25 eV) is apart from that of the bulk CB.  As can be seen in Fig. 2(b), the shape of the spin-resolved EDCs for the in-plane component is similar between $x$ = 1.0 and 0.9.  In addition, the spin polarization ($P_{\rm IP}$) [Fig. 2(c)] exhibits a similar value of $\sim$ 0.3, suggesting that the gapped SS in $x$ = 0.9 has a helical spin texture as in $x$ = 1.0.  Another notable aspect is that the in-plane spin polarization is gradually reduced upon approaching the QPT at $x$ $\sim$ 0.5, as evidenced by the smaller difference between the up- and down-spin EDCs for the samples with lower $x$ values.  The $P_{\rm IP}$ value in $x$ = 0.6 is $\sim$ 0.15 (half of that at $x$ = 1.0), and it vanishes in the non-topological phase ($x$ = 0.0).
 
 \begin{figure}[t]
\includegraphics[width=3.4in]{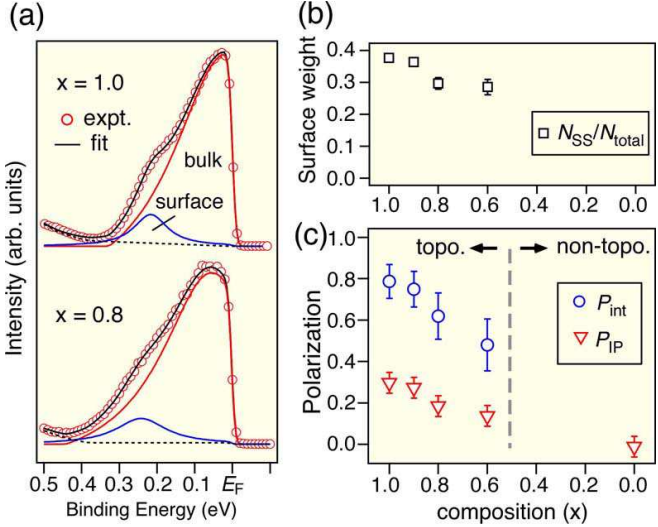}
\vspace{-0.4cm}
\caption{(Color online) (a) Representative results of the two-component fitting to the EDC obtained by the regular ARPES experiment to estimate the bulk and surface spectral weights as indicated by blue and red curves, respectively.  The plotted EDCs are generated by summing the raw EDCs within the {\bf k} window of the spin-resolved measurement, to take into account the {\bf k} broadening effect.  In the fitting, we assumed a Lorentzian peak for the surface, bulk VB, and bulk CB, multiplied by the Fermi-Dirac distribution function.  A constant term was subtracted from the peaks for bulk VB and CB to reproduce the bulk band gap.  (b) Ratio of the surface spectral weight to the total spectral weight, $N_{\rm SS}$/$N_{\rm total}$ (where $N_{\rm total}$ = $N_{\rm SS}$ + $N_{\rm bulk}$), at $E_{\rm SS}$ plotted against $x$ as estimated by the fitting of EDCs [see Fig. 3(a)].  (c) $P_{\rm IP}$ divided by $N_{\rm SS}$/$N_{\rm total}$, which represents the intrinsic spin polarization $P_{\rm int}$  of the SS (circles).  $P_{\rm IP}$ is also plotted with triangles.
}
\end{figure}

 \begin{figure}[t]
\includegraphics[width=3.4in]{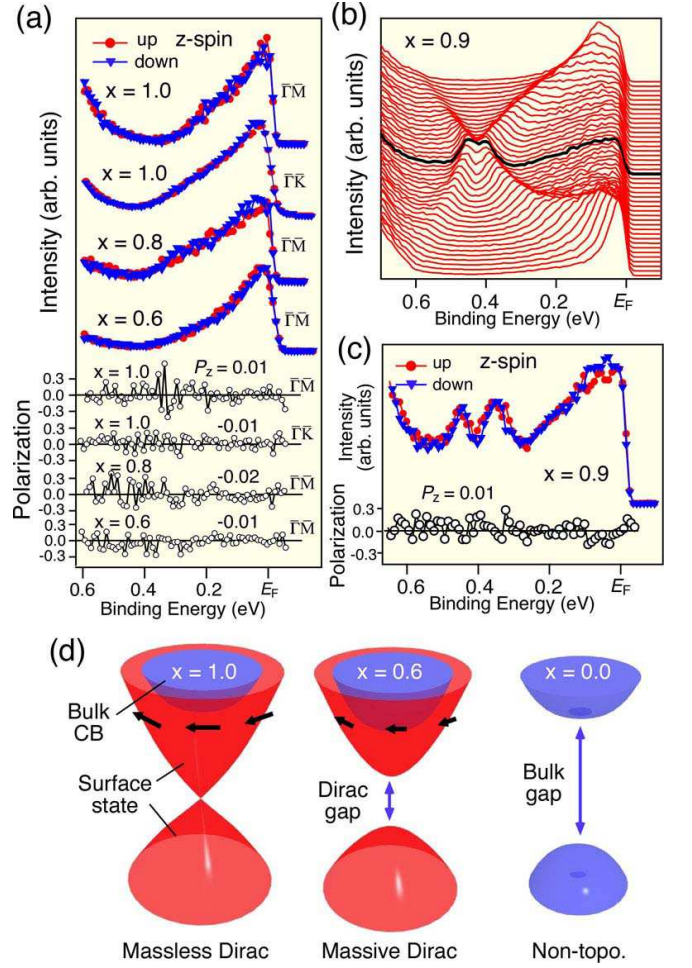}
\vspace{-0.4cm}
\caption{(Color online)(a) Spin-resolved EDCs and corresponding spin polarization for the out-of-plane component in the topological phase ($x$ = 1.0, 0.8, and 0.6) measured at the {\bf k} point along the $\bar{\Gamma}$$\bar{M}$ direction indicated by the red (thick) line in Fig. 2(a).  The EDCs along the $\bar{\Gamma}$$\bar{K}$ direction are also plotted for $x$ = 1.0.   (b) Near-$E_{\rm F}$ EDCs for $x$ = 0.9.  Spectrum at the $\bar{\Gamma}$ point is highlighted by thick black curve.  (c) Spin-resolved EDCs and corresponding spin polarization for the out-of-plane component measured at the $\bar{\Gamma}$ point for $x$ = 0.9.  (d) Schematic picture of the evolution of the topological SS (red) and bulk bands (blue) as a function of $x$.  The spin polarization vector is indicated by black arrows.
}
\end{figure}

 To discuss more quantitatively the absolute spin polarization of the SS, it is necessary to remove the finite contribution from the spin-degenerated bulk band.  For this purpose, we have fit the EDCs in the regular ARPES measurement by incorporating both the surface and bulk components [see Fig. 3(a)], and estimated the spectral weight of the SS ($N_{\rm SS}$) with respect to that of the total weight ($N_{\rm total}$ = $N_{\rm SS}$ + $N_{\rm bulk}$).  The intrinsic in-plane spin polarization of the SS ($P_{\rm int}$) corresponds to the $P_{\rm IP}$ value divided by $N_{\rm SS}$/$N_{\rm total}$.  (For the detailed procedure to estimate the spin polarization and corresponding error bars, see Supplemental Materials).  Figures 3(b) and (c) display the $x$-dependence of $N_{\rm SS}$/$N_{\rm total}$ and $P_{\rm int}$ ($P_{\rm IP}$ is also plotted for reference), respectively.  As can be seen in Fig. 3(c), the estimated $P_{\rm int}$ value, which reaches $\sim$ 0.8 for $x$ = 1.0, is gradually reduced with increasing $x$ to become $\sim$ 0.5 at $x$ = 0.6, and finally vanishes in the non-topological phase at $x$ = 0.0.  The origin of the reduction in the in-plane spin polarization toward the QPT in the gapped SS is not clear, but one can infer that it is not due to contamination from bulk conduction bands or the degenerate surface two-dimensional electron gas (2DEG) \cite{Hofmann}, since the effect of the former is already discounted in our analysis and the latter is obviously not formed in our surface situation.  The less-than-unity value of $P_{\rm int}$ at $x$ = 1.0 is likely to be a consequence of the spin-orbit entanglement as proposed from first-principle calculations \cite{Luie}.

The next important issue to clarify is the out-of-plane ($z$) spin component.  Figure 4(a) displays the spin-resolved EDCs and corresponding spin polarizations for the $z$-spin component measured at the same {\bf k} point as in Fig. 2 (along the $\bar{\Gamma}$$\bar{M}$ direction) for the samples in the topological phase ($x$ = 1.0, 0.8, and 0.6).  Spin-resolved EDCs for the {\bf k} point along the $\bar{\Gamma}$$\bar{K}$ direction are also displayed for $x$ = 1.0; obviously, there is no apparent $z$-spin component in the $\bar{\Gamma}$$\bar{M}$ nor the $\bar{\Gamma}$$\bar{K}$ directions.  It was previously found \cite{SoumaZspinPRL, MITzspin, HasanZspin} that in Bi$_2$Te$_3$ a finite $z$-spin texture emerges along the $\bar{\Gamma}$$\bar{K}$ direction due to strong hexagonal warping of the SS.  However, the hexagonal warping effect is much weaker in TlBi(S$_{1-x}$Se$_{x}$)$_{2}$ than in Bi$_2$Te$_3$, resulting in no or negligible $z$-spin component \cite{SoumaZspinPRL}.  The present data reconfirms this previous result for the whole topological phase, where the estimated $z$-axis spin polarization remains negligible (-0.02 $\sim$ +0.01).  

 While the above result demonstrates no $z$-spin polarization near the $k_{\rm F}$ point, a natural question arises as to whether any $z$-spin component exists exactly at the $\bar{\Gamma}$ point.  To address this issue, first we carefully examined the EDCs for $x$ = 0.9 [Fig. 4(b)] and determined the spectrum which exactly corresponds to the $\bar{\Gamma}$ point (thick line), and then performed the spin-resolved ARPES measurement at this {\bf k} point with a higher energy resolution (15 meV).  As shown in Fig. 4(c), we have succeeded in resolving two well-separated peaks in the spin-resolved EDCs attributed to the upper and lower branches of the gapped SS, respectively. Notably, there is essentially no difference between the up- and down-spin EDCs, indicating null $z$-spin polarization.  These results suggest that the $z$-spin component is absent irrespective of $x$ and {\bf k}.  Based on these observations, we depict in Fig. 4(d) the schematic picture of the spin-texture evolution for three different phases (massless Dirac, massive Dirac, and non-topological phases) in TlBi(S$_{1-x}$Se$_{x}$)$_{2}$.  

 Now we discuss the implications of the observed characteristics of the spin polarization in relation to the Dirac-gap opening.  First, the present result directly addresses the question whether the gapped SS has a topological origin.  One may argue that the topological phase transition should have occurred between the massless ($x$ = 1.0) and the massive ($x$ = 0.9) phase, and that the observed gapped SS corresponds simply to a pair of trivial, spin-degenerate SSs which happened to present a gap-like feature at the $\bar{\Gamma}$ point.  In this respect, such a possibility is immediately ruled out by the observation of the spin non-degenerate nature of the gapped SS.
 
  Another possible origin of the gapped SS is the hybridization of two topological SSs.  In ultrathin films of Bi$_2$Se$_3$, it was reported that a Dirac gap opens when the electron wavefunctions of the top and bottom surfaces start to hybridize \cite{XueNP, SQshenFilm}.  In this case, the hybridized state is spin degenerate and loses the original helical polarization \cite{TakayamaNanoLetter}. While the thick nature of our samples ($\sim$ 100 $\mu$m) already spoke against this possibility \cite{SatoNP}, the spin non-degeneracy observed here conclusively rules out the hybridization scenario.  Moreover, a recent theory for ultrathin films in the hybridized state predicted a finite $z$-spin component \cite{SQshenFilm}, which was not observed in the present experiment.  It is worth noting that recent ARPES studies on Bi$_2$Se$_3$ ultrathin films \cite{Berntsen} and bulk heterostructures involving Bi$_2$Se$_3$ units \cite{NakayamaHetero} have reported a second Dirac cone which originates from the backside surface (interface).  If the hybridization of the wave function between the Dirac cone at the topmost surface and the second Dirac cone at a backside surface (which may accidentally form due to bad cleaves) were the source of the observed phenomena, the separation between the topmost surface and the accidentally-formed backside surface must become systematically closer with decreasing $x$, which is extremely unlikely.
  
  An important question is whether the TRS is broken in the massive Dirac phase. When the topological SS feels a magnetic field or magnetization directed perpendicular to the surface (along the $z$-axis), it obtains a Dirac gap and the spin vector at the $\bar{\Gamma}$ point is aligned along the $z$-axis in opposite directions in the lower and upper branches of the gapped SS \cite{ME}.  In this respect, our finding that there is no $z$-spin component in the gapped SS does not give any support to this scenario.  Nevertheless, one should keep in mind that if there are magnetic domains with antiparallel polarizations, the overall $z$-spin component can be canceled. Hence, it is difficult to draw any firm conclusion about the TRS breaking from the present experiment.
  
  A viable possibility for the origin of the Dirac gap is the bulk-surface coupling.  A recent theory suggested that the Dirac-cone feature can be destroyed by strong potential impurities which lead to bulk-assisted scattering to allow virtual spin-flip excitations \cite{Balatzky}.  The proximity of the bulk band to the Dirac point near the QPT \cite{SatoNP} and the disorder due to random replacement of S with Se may be in line with this scenario.  Another possibility could be critical fluctuations associated with the QPT as mentioned in Ref. \cite{SatoNP}, and the reduction of the net spin polarization toward the QPT may be in line with this scenario; however, no evidence for such critical fluctuations has been obtained so far.  While a comprehensive answer to the origin of the Dirac gap has not yet been obtained, the present observation that established the nature of the spin texture and its evolution in the gapped phase provides an important step toward microscopic understanding of the topological QPT and the unconventional mass acquisition of Dirac fermions in TIs.
     
   In summary, we have reported our spin-resolved ARPES experiments on TlBi(S$_{1-x}$Se$_{x}$)$_{2}$ to determine the spin polarization as a function of $x$.  Besides firming up the intrinsic nature of the Dirac gap observed for 0.6 $\leq$ $x$ $\leq$ 0.9, we found that the helical spin texture survives in the massive Dirac phase, which clearly points to the topological nature of the gapped SS.  While the exact mechanism of the Dirac gap opening is yet to be elucidated, the direct connection between the emergence of the helical spin texture and the topological QPT, as well as the reduction of the net spin polarization toward to the QPT, gives an important clue.
   
\begin{acknowledgments}
We thank K. Kosaka and T. Arakane for their assistance in ARPES measurements, and T. Minami for his help in sample preparations.  This work was supported by JSPS (NEXT Program and KAKENHI 23224010), JST-CREST, MEXT of Japan (Innovative Area ``Topological Quantum Phenomena''), AFOSR (AOARD 104103 and 124038), and KEK-PF (Proposal number: 2012S2-001).
\end{acknowledgments}

\newpage
\onecolumngrid
\begin{center}
{\large Supplemental Materials for ``Spin Polarization of Gapped Dirac Surface States Near the 
Topological Phase Transition of TlBi(S$_{1-x}$Se$_{x}$)$_{2}$"}

\vspace{0.3 cm}

S. Souma,$^1$ M. Komastsu,$^2$ M. Nomura,$^2$ T. Sato,$^2$ A. Takayama,$^2$ T. Takahashi,$^{1,2}$
\newline
K. Eto,$^3$ Kouji Segawa,$^3$ and Yoichi Ando$^3$

{\footnotesize
$^1${\it WPI Research Center, Advanced Institute for Materials Research,
Tohoku University, Sendai 980-8577, Japan}
\newline
$^2${\it Department of Physics, Tohoku University, Sendai 980-8578, Japan}
\newline
$^3${\it Institute of Scientific and Industrial Research, Osaka University, Ibaraki, Osaka 567-0047, Japan}
}

\end{center}

\renewcommand{\thefigure}{S\arabic{figure}}
\setcounter{figure}{0}

\subsection{S1. Characteristics of the Dirac gap}

To elucidate the characteristics of the gapped surface state (SS) in more detail, we show in Fig. S1 the energy distribution curves (EDCs) at the $\bar{\Gamma}$ point for TlBi(S$_{1-x}$Se$_{x}$)$_{2}$  at each selenium composition $x$ in the topological phase.  As one can see in this plot, the EDC at the $\bar{\Gamma}$ point for $x$ = 1.0 is composed of a single peak at the Dirac point (DP) which is satisfactorily fitted by a single Lorentzian peak. On the other hand, we observe two well-separated peaks for $x$ = 0.9-0.6 (though the peaks for $x$ = 0.6 are broad), which are very well reproduced by two Lorentzian peaks. If the energy gap of the SS (the Dirac gap) were caused by a spectral-weight suppression at the DP as proposed in Ref.\cite{Xu_arXiv}, there is no reason that the experimental data presenting a two-peak feature (particularly the data points between the two peaks) are fitted well by a pair of Lorentzians. Our result thus suggests that the observed two-peak structure reflects an intrinsic spectral function of the SS with an energy gap at the DP.
 
 We note that the data for the gapped phase displayed in Fig. S1 were measured independently from our previous published work \cite{SatoNP} using different samples, and yet, the Dirac gap shows a quantitatively similar size indicative of its reproducibility. Specifically, the gap size in our previous experiment \cite{SatoNP}, as estimated from the peak position of the second-derivative of the EDCs, were $50\pm10, 70\pm10, 130\pm20$ meV for $x$ = 0.9, 0.8, and 0.6, respectively, and those in the present study as estimated from the energy separation of the peak position of two Lorentzians are $57\pm2, 73\pm2, 142\pm5$ meV, respectively, which agree well within the experimental uncertainties. In fact, as shown in Fig. S2 which plots the cleaving dependence of the band dispersion for $x$ = 0.9, the obtained gap feature is robust: the gap always exhibits the same value ($\sim55$ meV) despite the finite variations in the overall intensity distribution for different cleaves. This result is obviously incompatible with the recent proposal \cite{Xu_arXiv} that the ``gap'' can be caused by bad cleaves and rough surfaces, which would lead to sizable variation in the gap value for different cleaving conditions. It should also be noted that if the gap is caused by a momentum broadening, the spectral linewidth for the gapped sample should be broader than that for the samples showing no gap.  However, as demonstrated in Fig. S3, the peak width of the momentum distribution curves (MDCs) for $x$ = 0.9 is comparable to or even narrower than that for $x$ = 1.0, indicating that the momentum-broadening effect is unlikely to be the source of the gap opening.

One may argue that a ``gap'' could be caused by strong disorder and resultant smearing of the spectral weight at the DP.  If this is the case, it is naturally expected that a similar gap appears in other solid-solution systems of topological insulators.  In this respect, it is useful to compare the present result with Bi$_{2-x}$Sb$_{x}$Te$_{3-y}$Se$_{y}$ (BSTS) \cite{ArakaneBSTS} and Pb(Bi$_{1-x}$Sb$_x$)Te$_4$ (Pb124) \cite{SoumaPb124} where similar or even stronger random disorder is introduced. As one can see in Fig. S4, we found no evidence for a Dirac gap in both BSTS and Pb124 systems regardless of the ordered or disordered compositions, since the upper and lower branches of the Dirac cone are obviously degenerate exactly at the $\bar{\Gamma}$ point. Hence, in our experiments the Dirac gap has so far been only observed in the TlBi(S$_{1-x}$Se$_{x}$)$_{2}$ system in proximity to a topological QPT.

\subsection {S2. Procedure to estimate the spin polarization and its accuracy}

 In the analysis of the spin-resolved ARPES data, we first obtained the scattering asymmetry parameter $A = (N_{\rm L}-N_{\rm R})/(N_{\rm L}+N_{\rm R})$ at each binding energy where $N_{\rm L}$ and $N_{\rm R}$ are the number of electrons detected by the left-hand-side and right-hand-side channeltrons in the Mott spin polarimeter, respectively. Next we obtained the in-plane spin polarization $P_{\rm IP}$ through the formula $P_{\rm IP}=A/S_{\rm eff}$ where $S_{\rm eff}$ is the effective Sherman function ($S_{\rm eff}$ = 0.07; Ref.\cite{SoumaSpinRSI}). The statistical error in $P_{\rm IP}$ (${\Delta}P_{\rm IP}$) is directly linked to that in $A ({\Delta}A)$ through ${\Delta}P_{\rm IP} = {\Delta}A/S_{\rm eff}$. Since ${\Delta}A \approx {\Delta}N/N (N=N_{\rm L}+N_{\rm R})$, we accumulated all the spectra in Figs. 2(b) and (c) until the same error-bar range $(\pm5\%)$ is achieved for the $P_{\rm IP}$ value when it is averaged over the energy region of $\pm$0.1 eV centered at the surface-band energy. We plot this averaged $P_{\rm IP}$ value as a dashed line in Fig. 2(c) (with its actual value indicated), and also by triangles (with error bars of $\pm5\%$) in Fig. 3(c). The error bars in the intrinsic spin polarization $P_{\rm int}$ [$P_{\rm int} = P_{\rm IP}/(N_{\rm SS}/N_{\rm total})$] were evaluated from that in $P_{\rm IP}$ and $N_{\rm SS}/N_{\rm total}$ by taking into account the error propagation.

\subsection {REFERENCES}
\noindent
[15] T. Sato  {\it et al.}, Nature Phys. {\bf 7}, 840 (2011).\newline
[16] S. Souma {\it et al.}, Rev. Sci. Instrum. {\bf81}, 095101 (2010). \newline
[30] S.-Y. Xu {\it et al.}, arXiv:1206.0278. \newline
[31] T. Arakane  {\it et al.}, Nature Commun. {\bf 3}, 636 (2012). \newline
[32] S. Souma  {\it et al.}, Phys. Rev. Lett. {\bf 108}, 116801 (2012). \newline

 \begin{figure}[h]
 \includegraphics[width=3.4 in]{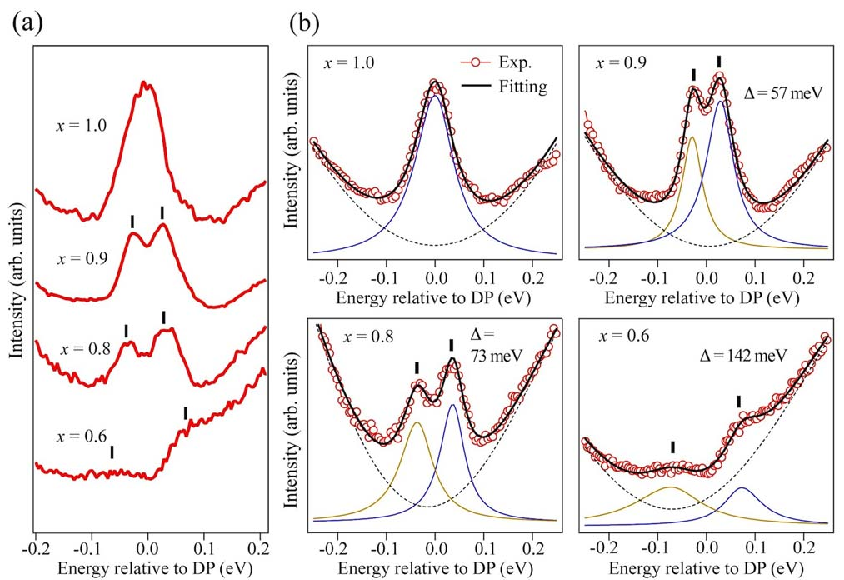}
\caption{(a) EDCs at the $\bar{\Gamma}$ point for various $x$ values measured in TlBi(S$_{1-x}$Se$_{x}$)$_{2}$. Black bar indicates the peak position of the Dirac-cone surface state estimated by the numerical fitting. (b) Results of fittings to the EDCs shown in (a). We assumed the Lorentzian peak(s) for the surface state (blue and yellow solid curves) and a polynomial function of up to the 4th order for the background (dashed curve). The size of the Dirac gap, as estimated from the energy difference between the peak positions of two Lorentzians, is also indicated for $x$ = 0.9, 0.8 and 0.6.
}
\end{figure}

\vspace{1 cm}

\begin{figure}[h]
\includegraphics[width=3.4 in]{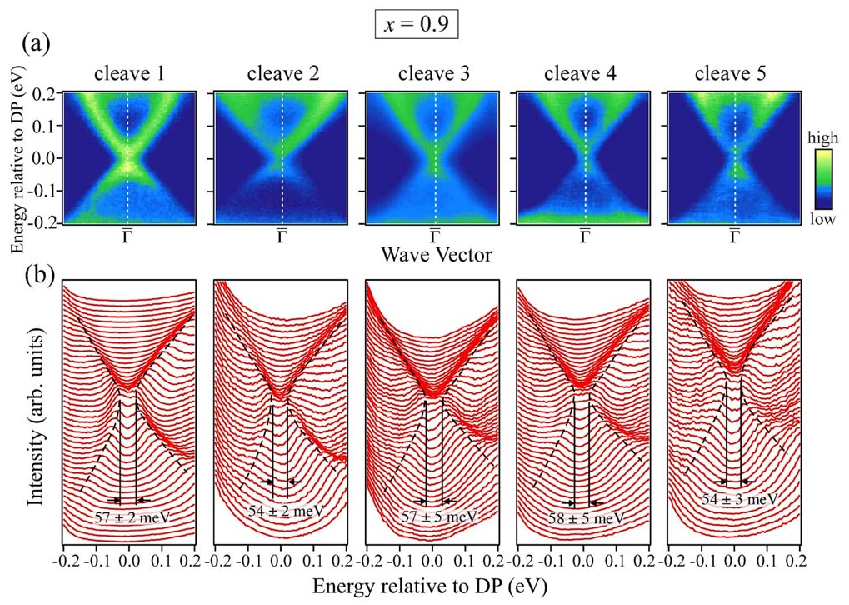}
\caption{(a) ARPES intensity plots around the DP of TlBi(S$_{0.1}$Se$_{0.9}$)$_2$ ($x$ = 0.9) for five different cleaved surfaces, and (b) the corresponding EDCs. Dashed curves in (b) are guides to the eyes to trace the surface-band dispersion.  The size of the Dirac gap, as estimated with the same procedure as in Fig. S1, is indicated by black solid lines.
}
\end{figure}

\begin{figure}
\includegraphics[width=3 in]{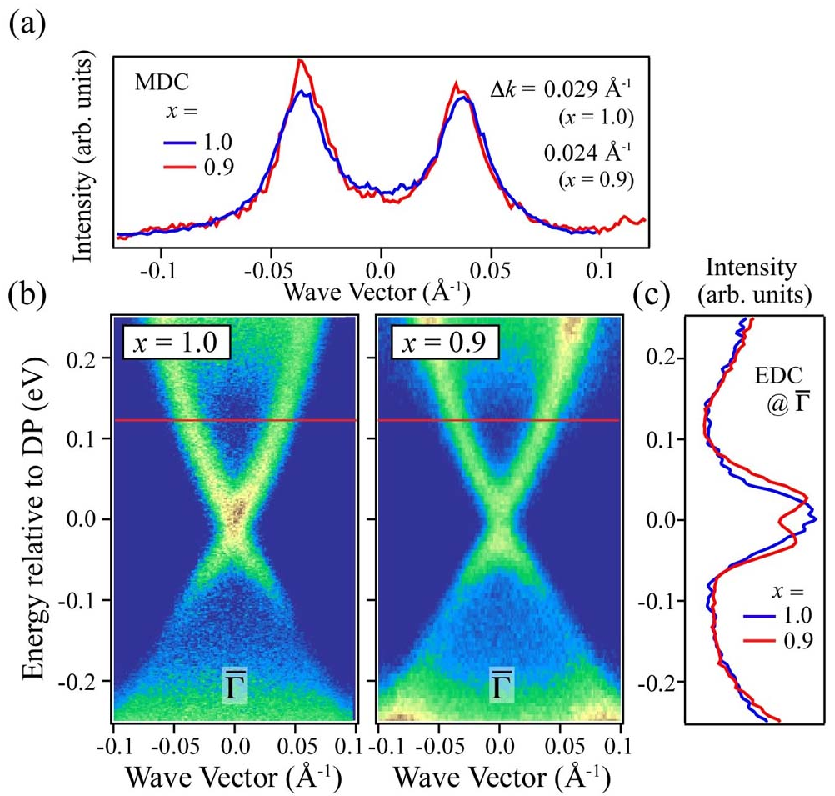}
\caption{Comparison of (a) the MDCs at 0.12 eV above the DP, (b) the ARPES intensities around the DP, and (c) the EDCs at the $\bar{\Gamma}$ point between $x$ = 1.0 and 0.9. The peak width of the MDC (${\Delta}k$) obtained by the fitting with Lorentzian peaks is also indicated in (a).
}
\end{figure}

\begin{figure}
\includegraphics[width=3.4 in]{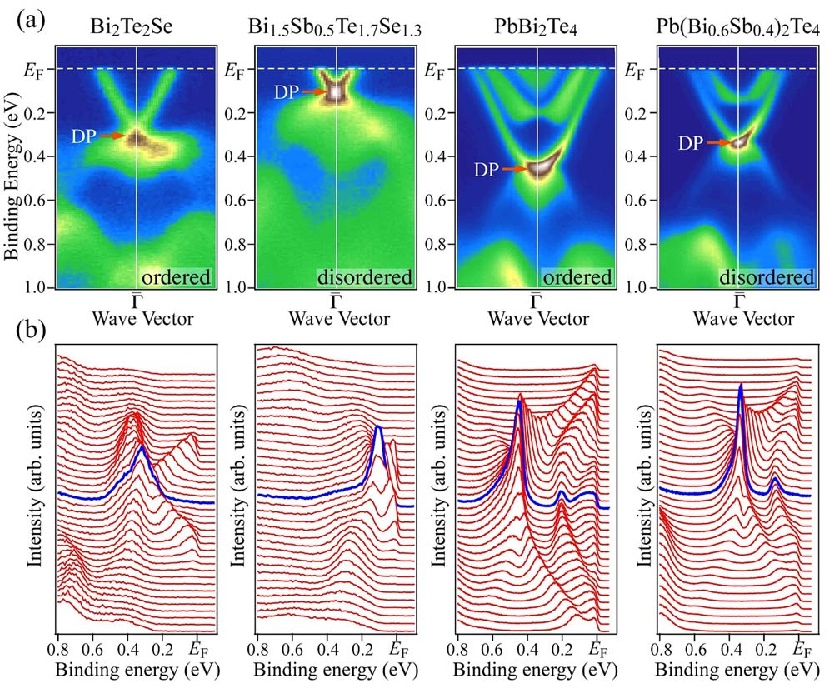}
\caption{(a) Near-$E_F$ ARPES intensity plots and (b) the corresponding EDCs for different material classes of topological insulators, Bi$_{2-x}$Sb$_{x}$Te$_{3-y}$Se$_{y}$ \cite{ArakaneBSTS} and Pb(Bi$_{1-x}$Sb$_x$)Te$_4$ \cite{SoumaPb124}. Comparisons between the ordered (Bi$_2$Te$_2$Se, PbBi$_2$Te$_4$) and disordered (Bi$_{1.5}$Sb$_{0.5}$Te$_{1.7}$Se$_{1.3}$, Pb(Bi$_{0.6}$Sb$_{0.4}$)$_2$Te$_4$) phases are presented. The EDCs at the $\bar{\Gamma}$ point are indicated by blue curves. None of them shows a Dirac gap.
}
\end{figure}

\vspace{5 cm}

    



\end{document}